\documentclass[11pt]{article}
\usepackage{amsmath}
\usepackage{graphicx}
\usepackage{cite}
\usepackage[left=3cm,right=3cm,top=3cm,bottom=2.5cm]{geometry}

\begin{document}

\title{Tagged-Particle Statistics in Single-File Motion with Random-Acceleration and Langevin Dynamics}
\author{Theodore  W. Burkhardt\\ Department of Physics, Temple University\\
Philadelphia, PA 19122, USA}

\maketitle

\begin{abstract}
In the simplest model of single-file diffusion, $N$ point particles wander on a segment of the $x$ axis of length $L$, with hard core interactions, which prevent passing, and with overdamped Brownian dynamics, $\lambda\dot{x}=\eta(t)$, where $\eta(t)$ has the form of Gaussian white noise with zero mean. In 1965 Harris showed that in the limit $N\to\infty$, $L\to\infty$ with constant $\rho=N/L$, the mean square displacement of a tagged particle grows subdiffusively, as $t^{1/2}$, for long times. Recently, it has been shown that the proportionality constants of the $t^{1/2}$ law  for randomly-distributed initial positions of the particles and for equally-spaced initial positions are not the same, but have ratio $\sqrt{2}$. In this paper we consider point particles on the $x$ axis, which collide elastically, and which move according to  (i) random-acceleration dynamics $\ddot{x}=\eta(t)$ and (ii) Langevin dynamics $\ddot{x}+\lambda\dot{x}=\eta(t)$. The mean square displacement and mean-square velocity of a tagged particle are analyzed for both types of dynamics and for random and equally-spaced initial positions and Gaussian-distributed initial velocities. We also study tagged particle statistics, for both types of dynamics, in the spreading of a compact cluster of particles, with all of the particles initially at the origin.
\end{abstract}

\maketitle

\newpage

\section{Introduction}\label{intro}

In the simplest model for single-file diffusion, $N$ point particles move on a segment of length $L$ of the $x$ axis, with hard core interactions between the particles, which prevent passing. Between collisions with its neighbors, each particle diffuses normally, with (overdamped) Brownian dynamics 
\begin{eqnarray}
&&\lambda\dot{x}=\eta(t)\,,\label{Browniandynamics}\\
&&\langle\eta(t)\rangle=0\,,\quad \langle\eta(t)\eta(t')\rangle=2\gamma\delta(t-t') \,,\label{eta}
\end{eqnarray}                                                                       
where $\lambda$ and $\gamma$ are constants, and the random force $\eta(t)$ has the form of Gaussian white noise with zero mean.  The best known characteristic of single-file diffusion, established by Harris \cite{Harris} a half century ago, is sub-diffusivity. In the limit $N\to\infty$, $L\to\infty$ with constant $\rho=N/L$, the mean square displacement of a tagged particle grows as $t^{1/2}$ in the long-time limit, in contrast to linear  $t$ dependence for non-interacting particles. Since the appearance of Harris's paper, single-file diffusion has been reexamined or studied theoretically in greater depth with many-different approaches  \cite{Jepsen,Levitt,Percus73,AlexanderPincus,vbkk,Arratia,Karger,rokaha,Aslangul,Kollmann,FlomembomTaloni,Kumar,LizanaAmbjornsson,Percus10,Lizanaetal,
Manzietal,LeibovichBarkai,Krapivskyetal,Lizanaetal2}. Recently, the proportionality constant in the $t^{1/2}$ power law has been shown \cite{Lizanaetal,Manzietal, LeibovichBarkai,Krapivskyetal} to depend on the initial configuration of the particles, with the value of the constant for an ``annealed" average over random initial positions of the particles greater by the factor $\sqrt{2}$ than for a ``quenched" initial configuration of equally-spaced particles.

Single-file diffusion has also been the subject of experimental studies. Introduced in the context of ion transport through cell membranes \cite{HodgkinKeynes}, it is also relevant to experiments on one-dimensional hopping, molecular motion in one-dimensional nanoporous materials, colloids in one-dimensional channels, and the sliding of proteins in a DNA sequence \cite{Richards,KargerRuthven,Chouetal,Meersmannetal,Lietal,Kuklaetal,Weietal,Lutzetal,Linetal,Dasetal,Siemsetal}. For an overview of single-file diffusion and an extensive list of theoretical and experimental papers, see the review by Ryabov \cite{Ryabov}. 

In this paper the model of hard point particles described above is studied, but with random-acceleration dynamics, 
\begin{equation} 
\ddot{x}=\eta(t)\,,\label{ranaccdynamics}
\end{equation}  
and the more general Langevin dynamics,
\begin{equation} 
\ddot{x}+\lambda\dot{x}=\eta(t)\,,\label{Langevindynamics}
\end{equation} 
which interpolates between (\ref{Browniandynamics}) and (\ref{ranaccdynamics}). 

We begin in Sect.~\ref{sect-homogeneousdist} with tagged-particle statistics in the single-file motion of an infinite number of point particles with homogeneous density. In Subsect.~\ref{subsect-Brownian} the results for Brownian dynamics we will need are reviewed. 
The case of particles which move with random-acceleration dynamics (\ref{ranaccdynamics}), collide elastically, and have Gaussian distributed initial velocities is considered in  Subsect.~\ref{subsect-ranacc}. The mean square displacement of a tagged particle is shown to increase as $t^{3/2}$ for long times, with different proportionality constants for random and equally-spaced initial particle positions, compared to $t^3$ for non-interacting, randomly-accerated particles.  In Subsect.~\ref{subsect-Langevin} a similar analysis is carried out for particles moving according to the Langevin dynamics (\ref{Langevindynamics}). The  derivation of the mean square displacement in Subsects.~\ref{subsect-ranacc} and \ref{subsect-Langevin} is simple and involves a mapping onto results for Brownian dynamics reviewed in Subsect.~\ref{subsect-Brownian}. For both random-acceleration and Langevin dynamics, the statistics of the velocity of a tagged particle is shown to be basically the same as for non-interacting particles. 

In Sect.~\ref{sect-compactinitialcluster} tagged particle statistics is studied in the spreading, through single-file motion, of a compact cluster of particles, with all of the particles initially located at the origin and with Gaussian distributed initial velocities. Aslangul's results \cite{Aslangul} for Brownian dynamics are reviewed in Subsect. \ref{subsect-Aslangul-Brownian}. In contrast to the subdiffusive behavior in Brownian systems with  homogeneous density $\rho$ described above, the mean displacement of a tagged particle in a cluster does not vanish, and the mean square displacement grows, for long times, as $t$ rather than $t^{1/2}$. Similar results have been found for Brownian particles with Gaussian-distributed initial particle positions in simulations \cite{FouadGawlinski}. In Subsects.~\ref{subsect-Aslangul-ranacc} and \ref{subsect-Aslangul-Langevin} analyses similar to Aslangul's are carried out for particles with random-acceleration and Langevin dynamics, respectively.   

Section \ref{ConcludingRemarks} contains concluding remarks.

\section{Tagged Particle Statistics, Homogeneous Particle Density}\label{sect-homogeneousdist}
\subsection{Brownian dynamics}\label{subsect-Brownian} 
  
For the Brownian dynamics (\ref{Browniandynamics}), the position at time $t$ of a non-interacting particle with initial positionly $x_0$ is given by 
\begin{equation}
x=x_0+\lambda^{-1}\int_0^t\,\eta(t')dt'\,.\label{x-Brownian}
\end{equation}
Averaging this equation and its square over the Gaussian white noise (\ref{eta}) yields   
\begin{equation}
\langle x\rangle=x_0\,,\quad\left\langle\left(x-\langle x\rangle\right)^2\right\rangle=2Dt\label{avs-Brownian}\,,
\end{equation} 
where $D=\gamma/\lambda^2$ is the diffusion coefficient.

The single-particle propagator or probability density $P_1(x, x_0;t)$ for propagation from $x$ to $x_0$ in a time $t$ satisfies the diffusion equation 
\begin{equation}
\left(\partial_t-D\,\partial_x^2\right)P_1(x, x_0;t)=0\label{fp-Brownian}
\end{equation} 
with initial condition $P_1(x,x_0;0)=\delta(x-x_0)$
and is given by
\begin{equation}
P_1\left(x,x_0;t\right)=(4\pi D t)^{-1/2}e^{-\left(x-x_0\right)^2/(4Dt)}\,,\label{prop1-Brownian}
\end{equation} 
consistent with the mean displacement and mean square displacement (\ref{avs-Brownian}).

For a system of $N$ hard point particles with positions $\vec{x}=(x_1,x_2,\dots,x_N)$ satisfying $x_1<x_2<\dots <x_N$ and initial positions  $\vec{x}_0=(x_{10},x_{20},\dots,x_{N0})$,  moving in single file with Brownian dynamics, the corresponding $N$-particle propagator satisfies the $N$-particle diffusion equation 
\begin{equation}
\left(\partial_t-D\sum_{i=1}^N\partial_{x_i}^2\right)P_N(\vec{x},\vec{x}_0;t)=0\label{fpN-Brownian}
\end{equation}
with initial condition
\begin{equation}
P_N(\vec{x},\vec{x}_0;0)=\prod_{i=1}^N \delta\left(x_i-x_{i0}\right)\label{ic-Brownian}
\end{equation}
and with the reflecting boundary condition \cite{rokaha}
\begin{equation}
\left(\partial_{x_i}-\partial_{x_{i+1}}\right)P_N(\vec{x},\vec{x}_0;t)\left\vert_{x_i=x_{i+1}}\right.=0\label{bc-Brownian}
\end{equation}
for $i=1,2,\dots,N-1$.
The solution is given by
\begin{eqnarray}
P_N(\vec{x},\vec{x'};t)&=&\sum_p\prod_{i=1}^N P_1\left(x_i,x_{i0}^p;t\right)\label{propN-Brownian1}\\
&=&\sum_p\prod_{i=1}^N (4\pi D t)^{-1/2}e^{-\left(x_i-x_{i0}^p\right)^2/(4D t)}\label{propN-Brownian2}
\end{eqnarray}
for $x_1<x_2<\dots <x_N$, where $P_1$ is the single-particle propagator in (\ref{prop1-Brownian}), and where $x_{i0}^p$  is the $i^{\rm th}$ element in the $p^{\rm th}$ permutation of the $N$ initial positions $\left(x_{10},x_{20},\dots, x_{N0}\right)$. 

The explicit form (\ref{propN-Brownian1}) of $P_N$ in terms of $P_1$ also follows directly from the equivalence of single-file motion and the motion of non-interacting particles on a line, which are free to pass each other, swapping
labels when they do. This equivalence, which is not limited to Brownian dynamics, is illustrated schematically for three particles in Fig. 1. The upper graph shows the trajectories of three particles undergoing single-file motion, which collide and reflect several times, with the particles arriving at $x_1,x_2,x_3$ originating at $x_{10},x_{20},x_{30}$, respectively.  In the lower graph the same curves are interpreted as trajectories of three non-interacting particles, which pass one another several times, with the particles arriving at $x_1,x_2,x_3$ originating at $x_{30},x_{10},x_{20}$. The right-hand side of equation (\ref{propN-Brownian1}) reflects this second interpretation and sums the probability density that the non-interacting particles arriving at $x_1,x_2,\dots,x_N$ 
originated at permutation $p$ of $(x_{10},x_{20},\dots,x_{N0})$ over the $N!$ distinct permutations.

Since any sum of single-particle terms of the form $\sum_{i=1}^N \langle{\cal F}(x_i)\rangle$, where ${\cal F}$ is an arbitrary function of position, is invariant under permutation of the particle labels, the sum, in the case of single-file motion, is the same as for non-interacting particles, which are free to pass each other. Together with the results (\ref{avs-Brownian}) for non-interacting particles, this implies 
\begin{equation}
\sum_{i=1}^N \langle x_i \rangle=\sum_{i=1}x_{i0}\,,\qquad\sum_{i=1}^N \langle x_i^2 \rangle=\sum_{i=1}^N\left(x_{i0}^2+2D t\right)\,.\label{sumrulex-Brownian}
\end{equation}

The mean square displacement of particle $n$ is obtained by multiplying the $N$-particle probability density (\ref{propN-Brownian2}) by $\left(x_n-x_{n0}\right)^2$ and integrating over the particle positions $x_1,\dots ,x_N$ with the constraint $x_1<x_2<\dots <x_N$. In the case of the annealed initial condition, one also integrates over a random distribution of the similarly constrained initial positions $x_{10},\dots ,x_{N0}$. For an  explicit calculation of the mean square displacement following this approach, see \cite{rokaha}. In the limit of an infinite number of particles, distributed randomly at $t=0$ with average density $\rho$, see e.g. \cite {Harris,Levitt,rokaha},     
\begin{equation}
\left\langle\left(x_n-x_{n0}\right)^2\right\rangle_{{\rm random}\;\vec{x}_0}
\approx{1\over\rho}\,\sqrt{4Dt\over\pi},\quad 2Dt\gg\rho^{-2}\,.\label{msd-Brownian-annealed}
\end{equation}
In the quenched case of particles with uniform initial spacing $\rho^{-1}$, analyzed in \cite{Lizanaetal,Manzietal, LeibovichBarkai,Krapivskyetal}, 
\begin{equation}
\left\langle\left(x_n-x_{n0}\right)^2\right\rangle_{{\rm equispace}\;\vec{x}_0}\approx{1\over\rho}\,\sqrt{2Dt\over\pi},\quad 2Dt\gg\rho^{-2}\,,\label{msd-Brownian-quenched}
\end{equation}
smaller by a factor $2^{-1/2}$ than (\ref{msd-Brownian-annealed}). Thus, the single-file restriction leads to sub-diffusive behavior. In contrast, for non-interacting Brownian particles,  $\langle\left(x_n-x_{n0}\right)^2\rangle=2Dt$, as shown in (\ref{avs-Brownian}). 

With macroscopic fluctuation theory, Krapivsky et al. \cite{Krapivskyetal} have recently confirmed the 2nd moments (\ref{msd-Brownian-annealed}) and  (\ref{msd-Brownian-quenched}), also calculated the 4th and 6th moments and the large deviation function for Brownian particles, and extended the analysis to other dynamical systems, such as the symmetric exclusion process.

\subsection{Random-Acceleration Dynamics}\label{subsect-ranacc}
We now consider the corresponding properties of particles which have random-acceleration dynamics (\ref{ranaccdynamics}), instead of Brownian dynamics, and which collide elastically.
The position $x$ and velocity $v$ of a non-interacting particle with equation of motion (\ref{ranaccdynamics}) evolve according to
\begin{eqnarray}
&&v(t)=v_0+\int_0^t\eta(t')\thinspace dt\,,\label{v-ranacc}\\
&&x(t)=x_0+v_0t+\int_0^t(t-t')\eta(t')\thinspace dt'.\label{x-ranacc}
\end{eqnarray}
Thus, $v(t)$ corresponds to a Brownian curve or random walk, and $x(t)$ to the integral of a Brownian curve.  Averaging these relations and their squares over the Gaussian white noise (\ref{eta}) yields
\begin{equation}
\begin{array}{l}\left\langle v\right\rangle=v_0\,,\\[2mm] \left\langle x\right\rangle=x_0+v_0t\,,\end{array}\quad\begin{array}{l}\left\langle\left(v-\langle v\rangle\right)^2\right\rangle=2\gamma t\,,\\[2mm] 
\left\langle\left(x-\langle x\rangle\right)^2\right\rangle={2\over 3}\gamma t^3\,.\end{array}\label{avs-ranacc}
\end{equation}

The one-particle propagator or probability density $P_1(x,v;x_0,v_0;t)$ for propagation from $\left(x_0,v_0\right)$ to $(x,v)$ in a time $t$
satisfies the Fokker-Planck equation \cite{Chandrasekhar,Risken}
\begin{equation}
\left({\partial\over\partial t}+v{\partial\over\partial x}-\gamma{\partial^2\over\partial v^2}\right)P_1(x,v;x_0,v_0;t) =0\label{fp-ranacc}
\end{equation}
with initial condition
\begin{equation}
P_1(x,v;x_0,v_0;0)=\delta(x-x_0)\delta(v-v_0)\label{initcond-ranacc}
\end{equation}
and is given explicitly by \cite{twb93}
\begin{eqnarray}
&&P_1(x,v;x_0,v_0;t)=3^{1/2}(2\pi\gamma t^2)^{-1}\nonumber\\
&&\quad\times\exp\left[-3(x-x_0-vt)(x-x_0-v_0t)/(\gamma t^3)-(v-v_0)^2/(\gamma t)\right].\label{prop1-ranacc}
\end{eqnarray}

The generalization of Eq. (\ref{prop1-ranacc}) in the case of $N$ randomly-accelerated particles on the $x$ axis is 
\begin{equation}
\left[{\partial\over\partial t}+\sum_{i=1}^N\left(v_i{\partial\over\partial x_i}-\gamma{\partial^2\over\partial v_i^2}\right)\right]P_N(\vec{x},\vec{v};\vec{x}_0,\vec{v}_0;t) =0\label{fpN-ranacc}\,,
\end{equation}
where $\vec{x}=(x_1,x_2,\dots,x_N)$, $\vec{v}=(v_1,v_2,\dots,v_N)$ . Since particles with equal masses simply exchange velocities in an elastic binary collision, the boundary condition for single-file motion is that $P_N$ be invariant under interchange of  $v_i$ and $v_{i+1}$ at each of the points $x_i=x_{i+1}$, where $i=1,2,\dots,N-1$. The solution of (\ref{fpN-ranacc}) which satisfies this boundary condition and reduces to $\prod_{i=1}^N\delta(x_i-x_{i0})\delta(v_i-v_{i0})$ at $t=0$ is
\begin{equation}
P_N(\vec{x},\vec{v};\vec{x}_0,\vec{v}_0;t)=\sum_p\prod_{i=1}^N P_1\left(x_i,v_i;x_{i0}^p,v_{i0}^p;t\right)\label{propN-ranacc1}
\end{equation}
for $x_1<x_2<\dots <x_N$, where $P_1$ is the single-particle propagator in Eq. (\ref{prop1-ranacc}). The quantity $x_{i0}^p$  in (\ref{propN-ranacc1}) is the $i^{\rm th}$ element in the $p^{\rm th}$ permutation of the initial positions $\left(x_{10},x_{20},\dots, x_{N0}\right)$, $v_{n0}^p$ is defined similarly, and $\sum_p$ indicates a sum over the $N!$ distinct permutations.  

Note the similarity between the $N$-particle propagator (\ref{propN-ranacc1}) for randomly-accelerated single-file dynamics and the corresponding  expression (\ref{propN-Brownian1}) for Brownian particles. Both follow directly from the equivalence of single-file motion with the motion of non-interacting particles on a line, which exchange labels whenever they pass each other. This is discussed just below (\ref{propN-Brownian2}) and illustrated in Fig. 1. 

An important point is that the equivalence between single-file motion and the motion of non-interacting particles which swap labels when they pass one another only holds for single-file motion with {\em elastic} collisions. When particle 1 passes particle 2, swapping labels, the velocities of particles 1 and 2 switch from $v_1$ to $v_2$ and from $v_2$ to $v_1$, respectively, conserving the quantities $v_1+v_2$ and $v_1^2+v_2^2$, just as if the particles had rebounded from each other elastically. 

Since any sum of single-particle terms of the form $\sum_{i=1}^N \langle{\cal F}(x_i,v_i)\rangle$, where ${\cal F}$ is an arbitrary function of $x_i$ and $v_i$, is invariant under permutation of the particle labels, the sum, in the case of single-file motion, is the same as for non-interacting particles moving on a line, which are free to pass each other. Below we will use the sum rules 
\begin{eqnarray}
&&\sum_{i=1}^N \langle v_i \rangle=\sum_{i=1}v_{i0}\,,\qquad\sum_{i=1}^N \langle v_i^2 \rangle=\sum_{i=1}^N\left(v_{i0}^2+2\gamma t\right)\,,\label{sumrulev}\\
&&\sum_{i=1}^N \langle x_i \rangle=\sum_{i=1}(x_{i0}+v_{i0}t)\,,\qquad\sum_{i=1}^N \langle x_i^2 \rangle=\sum_{i=1}^N\left[(x_{i0}+v_{i0}t)^2+\textstyle{{2\over 3}}\gamma t^3\right]\,,\label{sumrulex}
\end{eqnarray} 
where the right-hand side follows from the averages (\ref{avs-ranacc}) for non-interacting particle, or by explicit calculation utilizing the probability distribution (\ref{propN-ranacc1}).

Substituting (\ref{prop1-ranacc}) into (\ref{propN-ranacc1}) and integrating each of the velocities $v_1,\dots,v_N$ from $-\infty$ to $\infty$ leads to the probability distribution
\begin{equation}
Q_N(\vec{x};\vec{x}_0,\vec{v}_0;t)=\sum_p\prod_{i=1}^N \left(\textstyle{{4\over 3}}\pi\gamma t^3\right)^{-1/2}e^{-\left(x_i-x_{i0}^p-v_{i0}^p\,t\right)^2/\left({4\over 3}\gamma t^3\right)}\,,\label{posN-ranacc}
\end{equation}
for the positions of the particles at time $t$. 
Obtaining the mean square displacement is especially simple if (i) the particles are all initially at rest or (ii) the initial velocities are chosen from the Gaussian distribution  proportional to $e^{-\left(v_{10}^2+\dots+v_{N0}^2\right)/\left(2\sigma_{v_0}^2\right)}$. 
On averaging with this distribution of initial velocities, the probability distribution (\ref{posN-ranacc}) is replaced by 
\begin{equation}
\overline{Q}_N(\vec{x};\vec{x}_0;t)=\sum_p \prod_{i=1}^N \left[\pi\left(\textstyle{2\sigma_{v_0}^2 t^2+{4\over 3}}\gamma t^3\right)\right]^{-1/2}e^{-\left(x_i-x_{i0}^p\right)^2/\left(2\sigma_{v_0}^2 t^2+{4\over 3}\gamma t^3\right)}\,.\label{posN-ranacc-Gaussianv0} 
\end{equation} 
Below we only show results for Gaussian-distributed initial velocities. In the limit $\sigma_{v_0}^2\to 0$, the case of particles initially at rest is recovered.

Comparing  (\ref{propN-Brownian2}) and (\ref {posN-ranacc-Gaussianv0}), we see that randomly accelerated particles with Gaussian-distributed initial velocities are distributed along the $x$ axis just like Brownian particles, except that $2Dt$ is replaced by
\begin{equation}
2Dt\to\textstyle\sigma_{v_0}^2 t^2+{2\over 3}\,\gamma t^3\,.\label{replacements-ranacc}
\end{equation}
Making this replacement in (\ref{msd-Brownian-annealed}) and (\ref{msd-Brownian-quenched}), we obtain
\begin{eqnarray}
&&\left\langle\left(x_n-x_{n0}\right)^2\right\rangle_{\begin{subarray}{l}{\rm random}\;\vec{x}_0\\ {\rm Gauss}\,\vec{v}_0\end{subarray}}\approx\rho^{-1}\sqrt{{2\over\pi}\textstyle\left(\sigma_{v_0}^2 t^2+{2\over 3}\gamma t^3\right)}\,,
\label{msd-ranacc-annealed-Gaussianv0}\\
&&\left\langle\left(x_n-x_{n0}\right)^2\right\rangle_{\begin{subarray}{l}{\rm equispace}\;\vec{x}_0\\ {\rm Gauss}\,\vec{v}_0\end{subarray}}\approx\rho^{-1}\sqrt{{1\over \pi}\textstyle\left(\sigma_{v_0}^2 t^2+{2\over 3}\gamma t^3\right)}\,,
\label{msd-ranacc-quenched-Gaussianv0}
\end{eqnarray}
for $\sigma_{v_0}^2 t^2+{2\over3}\gamma t^3\gg\rho^{-2}$.

Just as for Brownian dynamics, the single-file restriction leads to anomalously slow growth of the mean square displacement. In contrast to (\ref{msd-ranacc-annealed-Gaussianv0}) and (\ref{msd-ranacc-quenched-Gaussianv0}), for non-interacting randomly-accelerated particles, 
 $\langle\left(x_n-x_{n0}\right)^2\rangle_{_{{\rm Gauss}\,\vec{v}_0}}=\sigma_{v_0}^2 t^2+{2\over 3}\,\gamma t^3$, as follows from (\ref{avs-ranacc}). 

In the limit $\gamma\to 0$ the large-$t$ behavior in (\ref{msd-ranacc-annealed-Gaussianv0}) and (\ref{msd-ranacc-quenched-Gaussianv0}) changes from $t^{3/2}$ to $t$, the same power law as for non-interacting Brownian particles, see (\ref{avs-Brownian}), with effective diffusion constant $D_{\rm eff}\propto \rho^{-1}\sigma_{v_0}$. In this limit the random acceleration is switched off, and the particles move with constant velocity between elastic collisions. The only source of randomness is the Gaussian distribution of the initial velocities.  

In the limit $\gamma\to 0$ the curves in Fig. 1 become straight lines. Nevertheless, the equivalence between single-file motion with elastic collisions and the motion of non-interacting particles which swap labels on passing one another continues to hold, as noted in the ``billiard-ball" section of Harris' 1965  paper \cite{Harris}. Expression (\ref{propN-ranacc1}) for the $N$-particle propagator still applies, but with single-particle propagator $P_1(x,v;x_0,v_0;t)=\delta(x-x_0-v_0t)\delta(v-v_0)$. The mean square displacement  (\ref{msd-ranacc-annealed-Gaussianv0}), in the special case $\gamma=0$, is derived in reference \cite{Harris}. 

We now turn to the tagged particle averages involving the velocity.  According to a calculation based on (\ref{propN-ranacc1}) and outlined in the Appendix,  in the limit of an infinite number of particles with homogeneous density $\rho$,
\begin{equation}  
\langle v_n\rangle_{_{{\rm Gauss}\,\vec{v}_0}}=0\,,\quad \langle v_n^2\rangle_{_{{\rm Gauss}\,\vec{v}_0}}=\sigma_{v_0}^2+ 2\gamma t\,.\label{avnandvnsqGauss}
\end{equation}
These results are the same as for non-interacting randomly-accelerated particles and consistent with the exact sum rules 
\begin{equation}
\sum_{i=1}^N \langle v_i \rangle_{_{{\rm Gauss}\,\vec{v}_0}}=0\,,\quad\sum_{i=1}^N \langle v_i^2 \rangle_{_{{\rm Gauss}\,\vec{v}_0}}=N\left(\sigma_{v_0}^2+2\gamma t\right)\,,\label{sumrulevsqGaussian}
\end{equation}
which follow from (\ref{sumrulev}) for Gaussian-distributed initial velocities. 

The mean square deviation of the velocity from its initial value, 
\begin{equation}
\langle\left(v_n-v_{n0}\right)^2\rangle_{_{{\rm Gauss}\,\vec{v}_0}}=\langle v_n^2\rangle_{_{{\rm Gauss}\,\vec{v}_0}}-2<v_nv_{n0}>_{_{{\rm Gauss}\,\vec{v}_0}}+\sigma_{v_0}^2\,,\label{meansqveldev}					
\end{equation}
is considerably more difficult to calculate from (\ref{propN-ranacc1}) than $\langle v_n^2\rangle_{_{{\rm Gauss}\,\vec{v}_0}}$.
For long times the first term on the right-hand side of (\ref{meansqveldev}), which  increases as $2\gamma t$, according to (\ref{avnandvnsqGauss}), is expected to dominate.\footnote{ For non-interacting particles, $<v_nv_{n0}>_{_{{\rm Gauss}\,\vec{v}_0}}=\sigma_{v_0}^2$, independent of $t$, as follows from (\ref{avs-ranacc}). In the case of single-file motion,  $<v_nv_{n0}>_{_{{\rm Gauss}\,\vec{v}_0}}$ equals $\sigma_{v_0}^2$ at $t=0$, but then decreases, with increasing $t$, since particles swap velocities in collisions, and the initial velocities of different particles are uncorrelated.} Thus, the no-passing restriction does not change the long-time behavior of $\langle\left(v_n-v_{n0}\right)^2\rangle$, even though it dramatically suppresses $\langle\left(x_n-x_{n0}\right)^2\rangle$.

\subsection{Langevin Dynamics}\label{subsect-Langevin}
We now consider the analog, for the Langevin dynamics (\ref{Langevindynamics}), of the above results for Brownian and random-acceleration dynamics. The position $x$ and velocity $v$ of a non-interacting particle with equation of motion (\ref{Langevindynamics}) evolve according to
\begin{eqnarray}
&&v(t)=v_0\,e^{-\lambda t}+\int_0^t e^{-\lambda(t-t')}\eta(t')dt'\,,\label{v-Langevin}\\
&&x(t)=x_0+\lambda^{-1}v_0\,\left(1-e^{-\lambda t}\right)+\lambda^{-1}\int_0^t\left(1-e^{-\lambda(t-t')}\right)\eta(t')dt'.\label{x-Langevin}
\end{eqnarray}
Averaging these relations and their squares over the Gaussian white noise (\ref{eta}) yields
\begin{equation}
\begin{array}{l}\left\langle v\right\rangle=v_0\,e^{-\lambda t}\,,\\[2mm] \left\langle x\right\rangle=x_0 + \lambda^{-1}v_0\left(1-e^{-\lambda t}\right)\,,\end{array}\quad\begin{array}{l}\left\langle\left(v-\langle v\rangle\right)^2\right\rangle=
\gamma\lambda^{-1}\,\left(1-e^{-2\lambda t}\right)\,,\\[2mm]
\left\langle\left(x-\langle x\rangle\right)^2\right\rangle=\gamma\lambda^{-3}\left(2\lambda t - 3 + 4e^{-\lambda t}-e^{-2\lambda t}\right)\,.\end{array}\label{avs-Langevin}
\end{equation}

The one-particle propagator or probability density $P_1(x,v;x_0,v_0;t)$ for propagation from $\left(x_0,v_0\right)$ to $(x,v)$ in a time $t$
satisfies the Chandrasekhar equation \cite{Chandrasekhar,Risken}
\begin{equation}
\left({\partial\over\partial t}+v{\partial\over\partial x}-\lambda-\lambda v{\partial\over\partial v}-\gamma{\partial^2\over\partial v^2}\right)P_1(x,v;x_0,v_0;t) =0\label{fp-Langevin}
\end{equation}
with initial condition
\begin{equation}
P_1(x,v;x_0,v_0;0)=\delta(x-x_0)\delta(v-v_0)\label{initcond}
\end{equation}
and is given explicitly by
\begin{eqnarray}
&&P_1(x,v;x_0,v_0;t)={1\over 2\pi\Delta^{1/2}}\exp\left[-\left(FS^2-2HRS+GR^2\right)/(2\Delta)\right]\,,\label{prop1-Langevin}\nonumber\\[2mm]
&&\quad R=x-x_0-v_0\,\lambda^{-1} \left(1-e^{-\lambda t}\right)\,,\nonumber\\
&&\quad S=v-v_0\, e^{-\lambda t}\,,\nonumber\\
&&\quad F=\gamma\lambda^{-3}\left(2\lambda t-3+4e^{-\lambda t}-e^{-2\lambda t}\right)\,,\nonumber\\
&&\quad G=\gamma\lambda^{-1}\left(1-e^{-2\lambda t}\right)\,,\nonumber\\
&&\quad H=\gamma\lambda^{-2}\left(1-e^{-\lambda t}\right)^2\,,\nonumber\\
&&\quad \Delta=FG-H^2\,.
\end{eqnarray}
The $N$-particle propagator has the same form (\ref{propN-ranacc1}) as in the preceding Subsection, but with the single-particle propagator (\ref{prop1-Langevin}). 

Substituting (\ref{prop1-Langevin}) into (\ref{propN-ranacc1}), integrating each of the velocities $v_1,\dots,v_N$ from $-\infty$ to $\infty$, and integrating the initial velocities over the Gaussian distribution introduced just above (\ref{posN-ranacc-Gaussianv0}) leads to the probability distribution
\begin{equation}
\bar{Q}_N(\vec{x};\vec{x}_0,\vec{v}_0;t)=\sum_p\prod_{i=1}^N \left(2\pi\sigma_{_{\rm Lan}}^2\right)^{-1/2}
e^{-\left(x_i-x_{i0}^p\right)^2/2\sigma_{_{\rm Lan}}^2}\,,\label{posN-Langevin-Gaussianv0}
\end{equation}
analogous to (\ref{posN-ranacc-Gaussianv0}), for the positions of the particles at time $t$,
where 
\begin{equation}
\sigma_{_{\rm Lan}}^2=\lambda^{-2}\left[\sigma_{v_0}^2\left(1-e^{-\lambda t}\right)^2+\gamma\lambda^{-1}\left(2\lambda t - 3 + 4e^{-\lambda t}-e^{-2\lambda t}\right)\right]\,.\label{sigmaLangevin}
\end{equation}

Comparing  (\ref{propN-Brownian2}) and (\ref{posN-Langevin-Gaussianv0}), 
we see that randomly-accelerated particles with Gaussian-distributed initial velocities are distributed along the $x$ axis just like Brownian particles, except that $2Dt$ is replaced by
\begin{equation}
2Dt\to\sigma^2_{\rm Lan}\,.\label{replacements-Langevin}
\end{equation}
Making  this replacement in (\ref{msd-Brownian-annealed}) and (\ref{msd-Brownian-quenched}) yields
\begin{eqnarray}
&&\left\langle\left(x_n-x_{n0}\right)^2\right\rangle_{\begin{subarray}{l}{\rm random}\;\vec{x}_0\;\;\\ {\rm Gauss}\,\vec{v}_0\end{subarray}}\;
\approx{1\over\rho}\,\sqrt{{2\over\pi}\,\sigma_{_{\rm Lan}}^2}\;,\label{msd-Langevin-annealed-Gaussianv0}\\
&&\left\langle\left(x_n-x_{n0}\right)^2\right\rangle_{\begin{subarray}{l}{\rm equispace}\;\vec{x}_0\\ {\rm Gauss}\,\vec{v}_0\end{subarray}}\;\approx{1\over\rho}\,\sqrt{{1\over\pi}\,\sigma_{_{\rm Lan}}^2}\;,\label{msd-Langevin-quenched-Gaussianv0}
\end{eqnarray}
where (\ref{msd-Langevin-annealed-Gaussianv0}) and (\ref{msd-Langevin-quenched-Gaussianv0})
hold asymptotically for 
$\sigma_{_{\rm Lan}}^2\gg\rho^{-2}$. In contrast, for non-interacting particles with Langevin dynamics,  $\langle\left(x_n-x_{n0}\right)^2\rangle_{_{{\rm Gauss}\,\vec{v}_0}}=\sigma_{_{\rm Lan}}^2$, as follows from (\ref{avs-Langevin}) and (\ref{sigmaLangevin}). 

The quantity $\sigma_{_{\rm Lan}}^2$ in (\ref{sigmaLangevin}) is a monotonically increasing function of $t$ with the asymptotic behavior
\begin{equation}
\sigma_{_{\rm Lan}}^2\approx\left\{\begin{array}{l}\,\sigma_{v_0}^2 t^2+{2\over 3}\gamma t^3\,,\\ \sigma_{v_0}^2\lambda^{-2}+2\gamma\lambda^{-2}t\,,\end{array}\right.\begin{array}{l}0<\lambda t\ll 1\,,\\ \lambda t\gg 1\,.\end{array}\label{sigmaLangevinasymptotes}
\end{equation}
Substituting these asymptotic forms in (\ref{posN-Langevin-Gaussianv0}) and comparing with (\ref{propN-Brownian2}) and (\ref{posN-ranacc-Gaussianv0}), one finds that the distribution function (\ref{posN-Langevin-Gaussianv0}) for the positions of particles with Langevin dynamics reduces to the corresponding distributions for random-acceleration dynamics and Brownian dynamics in the short and long-time limits, respectively. The diffusion coefficient in the Brownian limit is $D=\gamma/\lambda^2$.

For Langevin dynamics the analogs of the velocity averages (\ref{avnandvnsqGauss})
are 
\begin{equation}  
\langle v_n\rangle_{_{{\rm Gauss}\,\vec{v}_0}}=0\,,\quad \langle v_n^2\rangle_{_{{\rm Gauss}\,\vec{v}_0}}= \sigma_{v_0}^2 e^{-2\lambda t}+\gamma\lambda^{-1}\,\left(1-e^{-2\lambda t}\right)\,,\label{avnandvnsqLangevin}
\end{equation}
the same as for non-interacting particles in (\ref{avs-Langevin}). As $t$ increases from $t=0$ to $\infty$, $\langle v_n^2\rangle_{_{{\rm Gauss}}}$ changes monotonically from  $\sigma_{v_0}^2$ to $\gamma\lambda^{-1}$. If the system of particles is initially in thermal equilibrium, ${1\over 2}m\sigma_{v_0}^2={1\over 2}k_BT$, and if ${1\over 2}m\gamma\lambda^{-1}={1\over 2}k_BT$, it approaches thermal equilibrium in the long-time limit. If both of these conditions are fulfilled, the system never leaves equilibrium.

\section{Tagged Particle Statistics in a Compact Cluster}\label{sect-compactinitialcluster}
Having thus far considered systems with homogeneous density $\rho$, we now turn to the contrasting case of particles which form a compact cluster. 
Aslangul \cite{Aslangul} has analyzed tagged particle statistics in an expanding cluster of $N$ Brownian particles, with all the particles initially at the origin. In this Section we review some of his findings and derive analogous results for random-acceleration and Langevin dynamics. 

\subsection{Brownian dynamics}\label{subsect-Aslangul-Brownian}
On setting $x_{10}=x_{20}=\dots=x_{N0}=0$ in (\ref{propN-Brownian1}) and (\ref{propN-Brownian2}), all $N!$ terms in the sum over permutations become identical, so that 
\begin{equation}
P_N(\vec{x},\vec{0};t)=N!\prod_{i=1}^N P_1\left(x_i,0;t\right)\,,\quad P_1\left(x,0;t\right)=(4\pi D t)^{-1/2}e^{-x^2/4Dt}\,.\label{propN-Brownian3}
\end{equation}
The probability density $P^{(n)}(x_n,t)$ of tagged particle $n$ is obtained by integrating (\ref{propN-Brownian3}) over all of the $x_1,\dots, x_N$  except $x_n$, with constraint $-\infty<x_1<x_2<\dots x_N<\infty$. This yields
 \begin{equation}
P^{(n)}(x,t)={N!\over(n-1)!(N-n)!}(4\pi D t)^{-1/2}e^{-x^2/4Dt}f(x)^{n-1}f(-x)^{N-n} \,,\label{aslangul-dist-Brownian}
\end{equation}
where we omit the subscript on $x_n$ for simplicity, and where 
\begin{equation}
f(x)=\int_{-\infty}^x P_1\left(x',0;t\right)dx'=\textstyle{1\over 2}\,\left[1+{\rm erf}\left({x\over\sqrt{4Dt}}\right)\right]\,.\label{aslangul-f(x)-Brownian}
\end{equation}

It is straightforward to determine the mean displacement $\langle x_n\rangle$ and  mean square displacement $\langle x_n^2\rangle$ of particle $n$ for the distribution  (\ref{aslangul-dist-Brownian}) by numerical integration. For $N\gg1$,  Aslangul obtained the analytical predictions
\begin{eqnarray}
&&\langle x_N\rangle=-\langle x_1\rangle\approx\sqrt{\,\left(4\ln{N\over\sqrt{4\pi}}\right)Dt\,}\,\,,\label{<x_N>-Brownian}\\
&&\left\langle\left(x_N-\langle x_N\rangle\right)^2\right\rangle=\left\langle\left(x_1-\langle x_1\rangle\right)^2\right\rangle\approx
{e^{2/3}\over (2\pi)^{1/3}\ln{N\over\sqrt{4\pi}}}Dt\,,\label{<x_N^2>cum-Brownian}
\end{eqnarray}
for the first and last particles in a file of $N$ particles,  
and 
\begin{equation}
\langle x_{_{(N+1)/2}}\rangle=0\,,\qquad
\left\langle x_{_{(N+1)/2}}^2\right\rangle\approx{\pi\over N}\,Dt\,.\label{<x_(N+1)/2>cum-Brownian}
 \end{equation}
for the middle particle for $N$ odd. 

A useful result, not given in \cite{Aslangul}, is the steepest-descent prediction for $N\gg1$, $n\gg 1$, and $N-n\gg 1$, i.e., for particles well away from the ends of a file of a large number of particles:  
\begin{eqnarray}
&&\langle x_n\rangle \approx\left[2\,{\rm erf}^{-1}(2\nu-1)\right]\sqrt{Dt}\,,\label{<xn>-Brownian}\\
&&\left\langle\left(x_n-\langle x_n\rangle\right)^2\right\rangle\approx{4\pi\over N}\,\nu(1-\nu)
\exp\left(2\left[{\rm erf}^{-1}(2\nu-1)\right]^2\right)Dt\,,\label{<xnsq>cum-Brownian}\\
&&\nu\equiv{n+\alpha\over N+1+2\alpha}\,.\label{nudef}
\end{eqnarray}
Here $\alpha$ is a dimensionless constant of order of magnitude 1, so that $\nu\approx n/N$ to leading order for $N$ large. We have included $\alpha$ in the definition (\ref{nudef}) of $\nu$ to ensure that (\ref{<xn>-Brownian})-(\ref{nudef}) are consistent with the exact symmetry properties $\langle x_n\rangle=-\langle x_{N-n+1}\rangle$ and 
$\langle x_n^2\rangle=\langle x^2_{N-n+1}\rangle$ for the $n$th particles from the left end and from the right end of the file. Making the replacement
$\sum_n\to N\int d\nu$ and utilizing $\int_0^1\left[{\rm erf}(2\nu-1)\right]^2d\nu={1\over 2}$, it is simple to check the consistency of (\ref{<xn>-Brownian})-(\ref{nudef}), for large $N$, with the sum rules (\ref{sumrulex-Brownian}). 

Comparing (\ref{<x_N>-Brownian})-(\ref{nudef}) with the corresponding results, $\langle x_n\rangle=0$ and $\langle\left(x_n-\langle x_n\rangle\right)^2\rangle=2Dt$, for non-interacting particles, we see that in the single-file diffusion of $N$ particles which are initially located at the origin,
\begin{enumerate}
\item $\langle x_n\rangle$ does not vanish, since the particles migrate toward regions of lower density, away from the origin, but varies as $t^{1/2}$,
\item in both single-file and ordinary diffusion, $\langle\left(x_n-\langle x_n\rangle\right)^2\rangle$ increases linearly with $t$, but in the case of single-file diffusion the proportionality constant is suppressed by an $n$-dependent prefactor proportional to  $N^{-1}$, which crosses over to $(\ln N)^{-1}$ near the ends of the file.
\end{enumerate}

In Figs. 2 and 3 essentially exact results for $\langle x_n\rangle$ and $\langle\left(x_n-\langle x_n\rangle\right)^2\rangle$, obtained from (\ref{aslangul-dist-Brownian}) by numerical integration and indicated by black points, are compared with the steepest-descent predictions (\ref{<xn>-Brownian})-(\ref{nudef}) for $N=501$ particles and for $\alpha=1$ and $\alpha=-1$. There is excellent agreement for all the particles except those near the ends of the file. On adjusting $\alpha$ to reproduce the rightmost point in Fig. 3, excellent agreement is achieved for all 501 particles. 

\subsection{Random acceleration dynamics}\label{subsect-Aslangul-ranacc}
Proceeding as in Eqs. (\ref{propN-Brownian3})-(\ref{aslangul-f(x)-Brownian}), but with the random-acceleration propagators (\ref{prop1-ranacc}) and (\ref{posN-ranacc-Gaussianv0}) instead of the Brownian expressions, leads to the probability density 
\begin{equation}
\bar{P}^{(n)}(x,v;t)={N!\over(n-1)!(N-n)!}\, \bar{P}_1\left(x,v;t\right) \, \bar{f}(x)^{n-1}\bar{f}(-x)^{N-n} \label{aslangul-dist-ranacc}
\end{equation}
for particle $n$, where
\begin{eqnarray}
&& \bar{P}_1\left(x,v;t\right)={1\over\sqrt{2\pi \sigma_{v_0}^2}}\int_{-\infty}^\infty e^{-v_0^2/(2\sigma_{v_0}^2)}P_1(x,v;0,v_0;t)\,dv_0\nonumber\\
&&\qquad\qquad={1\over\sqrt{\pi\left(2\sigma_{v_0}^2 t^2+{4\over 3}\gamma t^3\right)}}\,\exp\left(-{x^2\over 2\sigma_{v_0}^2 t^2+{4\over 3}\gamma t^3}\right)
\nonumber\\[3mm] 
&&\qquad\qquad\times\;\sqrt{{3\over 4\pi\gamma t}\,{\sigma_{v_0}^2+{2\over 3}\gamma t\over \sigma_{v_0}^2+{1\over 2}\gamma t}}\, 
\exp\left[-{3\over 4\gamma t}\,{\sigma_{v_0}^2+{2\over 3}\gamma t\over \sigma_{v_0}^2+{1\over 2}\gamma t} 
\left(v-{x\over t}\,{\sigma_{v_0}^2+\gamma t\over\sigma_{v_0}^2+{2\over 3}\gamma t }\right)^2\right]\label{aslangul-ranacc-P1bar}
\end{eqnarray}
is a product of two normalized Gaussian distributions, and where
\begin{equation}
\bar{f}(x)=\int_{-\infty}^x dx'\int_{-\infty}^\infty dv\,\bar{P}_1(x',v;t)=\textstyle{1\over 2}\,\left[1+{\rm erf}\left({x\over\sqrt{2\sigma_{v_0}^2 t^2+{4\over3}\gamma t^3
}}\right)\right]\,.\label{aslangul-ranacc-fbar}
\end{equation}

Integrating (\ref{aslangul-dist-ranacc}) over all $v$, using
\begin{equation}
\int_{-\infty}^\infty\bar{P}_1(x,v;t)\, dv={e^{-x^2/\left(2\sigma_{v_0}^2 t^2+{4\over 3}\gamma t^3\right)}\over\sqrt{\pi\left(2\sigma_{v_0}^2 t^2+{4\over 3}\gamma t^3\right)}}\,,
\end{equation}
and comparing with (\ref{aslangul-dist-Brownian}) and (\ref{aslangul-f(x)-Brownian}),
we find that the positions of the $N$ randomly-accelerated particles which are initially all at the origin with Gaussian-distributed initial velocities, are distributed  just as in the Brownian case, except that $2Dt$ is replaced by $\sigma_{v_0}^2 t^2+{2\over 3}\gamma t^3$, the same prescription (\ref{replacements-ranacc}) already encountered in Sect.~\ref{sect-homogeneousdist}. With this replacement, all of the Brownian results 
(\ref{<x_N>-Brownian})-(\ref{nudef}) for the moments of the tagged-particle position $x_n$ apply to randomly-accelerated particles. Thus, the mean position 
$\langle x_n\rangle_{{\rm Gauss\;\vec{v}_0}}$ is proportional  $\left(\sigma_{v_0}^2 t^2+{2\over 3}\gamma t^3\right)^{1/2}$, and the variance $\langle\left(x_n-\langle x_n\rangle\right)^2\rangle_{{\rm Gauss\;\vec{v}_0}}$ increases as $\sigma_{v_0}^2 t^2+{2\over 3}\gamma t^3$, suppressed by a prefactor proportional to $N^{-1}$, which crosses over to $(\ln N)^{-1}$ near the ends of the file. In contrast, or particles which are free to pass each other, $\langle\left(x_n-\langle x_n\rangle\right)^2\rangle_{{\rm Gauss\;\vec{v}_0}}=\sigma_{v_0}^2 t^2+{2\over 3}\gamma t^3$, as follows from (\ref{avs-ranacc}).

Equations (\ref{aslangul-dist-ranacc})-(\ref{aslangul-ranacc-fbar}) determine the average of any function of $x_n$ and $v_n$. In addition to the results for $\langle x_n\rangle_{{\rm Gauss\;\vec{v}_0}}$ and $\langle x_n^2\rangle_{{\rm Gauss\;\vec{v}_0}}$ given in the preceding paragraph, one finds 
\begin{eqnarray}
&&\langle v_n\rangle_{_{\rm Gauss}\,\vec{v}_0}={1\over t}\,{\sigma_{v_0}^2+\gamma t\over\sigma_{v_0}^2+{2\over 3}\gamma t }\,\langle x_n\rangle_{{\rm Gauss}\,\vec{v}_0}\,,\label{avvn-Aslangul-ranacc}\\[2mm]
&&\left\langle  v_n^2\right\rangle_{{\rm Gauss}\,\vec{v}_0}=\left({1\over t}\,{\sigma_{v_0}^2+\gamma t\over\sigma_{v_0}^2+{2\over 3}\gamma t }  \right)^2\left\langle x_n^2\right\rangle_{_{\rm Gauss}\,\vec{v}_0}+{\textstyle{2\over 3}}\,\gamma t\,{\sigma_{v_0}^2+{1\over 2}\gamma t\over \sigma_{v_0}^2+{2\over 3}\gamma t}\,.\label{avvnsq-Aslangul-ranacc}
\end{eqnarray}

Since $\langle x_n\rangle_{{\rm Gauss}\;\vec{v_0}}\propto\left(\sigma_{v_0}^2 t^2+{2\over3}\gamma t^3\right)^{1/2}$, as follows from  (\ref{<xn>-Brownian}) with replacement (\ref{replacements-ranacc}), the result (\ref{avvn-Aslangul-ranacc}) is consistent with $\langle v_n\rangle=d\langle x_n\rangle/dt$, just as in the case of non-interacting particles (see (\ref{avs-ranacc})). We note that  (\ref{avvnsq-Aslangul-ranacc})
is fully consistent with the exact sum rules (\ref{sumrulevsqGaussian}) and
\begin{equation} 
\sum_{i=1}^N\langle x_i^2\rangle_{_{{\rm Gauss}\,\vec{v}_0}}=\textstyle N\left(\sigma_{v_0}^2 t^2+{2\over 3}\gamma t^3\right)\label{sumrulexsqGaussian}\,,
\end{equation}
which follows from (\ref{sumrulex}) for particles initially at the origin with Gaussian-distributed initial velocities. 

\subsection{Langevin dynamics}\label{subsect-Aslangul-Langevin}
Proceeding as in the preceding Subsection, but with $Q_N(\vec{x};\vec{x}_0,\vec{v}_0;t)$ given by (\ref{posN-Langevin-Gaussianv0}) instead of (\ref{posN-ranacc-Gaussianv0}), we find that with the replacement $2Dt\to\sigma^2_{\rm Lan}$, already encountered in (\ref{sigmaLangevin}) and (\ref{replacements-Langevin}),
all of the Brownian results (\ref{<x_N>-Brownian})-(\ref{nudef}) for the moments of the tagged-particle position $x_n$ apply to particles with Langevin dynamics. 

These and other averages involving $x_n$ and $v_n$ may be calculated from the probability density $\bar{P}^{(n)}(x,v;t)$, which is the same as in (\ref{aslangul-dist-ranacc}), except that  
\begin{eqnarray}
&&\bar{P}_1\left(x,v;t\right)={1\over\sqrt{2\pi\sigma_{\rm Lan}^2}}\,\exp\left(-{x^2\over 2\sigma_{\rm Lan}^2}\right)\;
{1\over\sqrt{2\pi h(t)^2}}\exp\left[-{\left(v-g(t)\,x\right)^2\over 2h(t)^2}\right]\,,\label{aslangul-Langevin-P1bar}\\
&&g(t)=\partial_t\ln\sigma_{_{\rm Lan}}={1\over\lambda\sigma_{_{\rm Lan}}^2}\left(1-e^{-\lambda t}\right)\left[\sigma_{v_0}^2\, e^{-\lambda t}+\gamma\lambda^{-1}\left(1-e^{-\lambda t}\right)\right]\,,\label{aslangul-Langevin-g}\\[2mm]
&&h(t)^2=\sigma_{v_0}^2 e^{-2\lambda t}+\gamma\lambda^{-1}\left(1-e^{-2\lambda t}\right)-g(t)^2\sigma_{_{\rm Lan}}^2\,,\label{aslangul-Langevin-h}\\[2mm]
&&\bar{f}(x)=\textstyle{1\over 2}\,\left[1+{\rm erf}\left({x\over\sqrt{2\sigma_{_{\rm Lan}}^2}}\right)\right]\,.\label{aslangul-Langevin-fbar}
\end{eqnarray}
The first two moments of $v_n$, the analogs of (\ref{avvn-Aslangul-ranacc}) and (\ref{avvnsq-Aslangul-ranacc}), are
\begin{eqnarray}
&&\langle v_n\rangle_{_{\rm Gauss}\,\vec{v}_0}=g(t)\,\langle x_n\rangle_{{\rm Gauss}\,\vec{v}_0}\,,\label{avvn-Aslangul-Langevin}\\
&&\left\langle  v_n^2\right\rangle_{{\rm Gauss}\,\vec{v}_0}=g(t)^2\,\left\langle x_n^2\right\rangle_{_{\rm Gauss}\,\vec{v}_0}+h(t)^2\,.\label{avvnsq-Aslangul-Langevin}
\end{eqnarray}
Since $\langle x_n\rangle_{{\rm Gauss}\;\vec{v_0}}\propto\sigma_{\rm Lan}$, as follows from  (\ref{<xn>-Brownian}) with replacement $2Dt\to\sigma^2_{\rm Lan}$, the result (\ref{avvn-Aslangul-Langevin}), is consistent with $\langle v_n\rangle=d\langle x_n\rangle/dt$, just as in the case of non-interacting particles (see (\ref{avs-Langevin})). We note that (\ref{avvnsq-Aslangul-Langevin}) is consistent with the  
sum rules
\begin{eqnarray} 
&&\sum_{i=1}^N\langle v_i^2\rangle_{_{{\rm Gauss}\,\vec{v}_0}}=N\left[\sigma_{v_0}^2+\gamma\lambda^{-1} \left(1-e^{-2\lambda t}\right)\right]\,,\label{sumrulevsqLangevin}\\
&&\sum_{i=1}^N\langle x_i^2\rangle_{_{{\rm Gauss}\,\vec{v}_0}}=N \sigma_{\rm Lan}^2\,,\label{sumrulexsqLangevin}
\end{eqnarray}
which are the analogs for Langevin dynamics of  (\ref{sumrulevsqGaussian}) and (\ref{sumrulexsqGaussian}).

\section{Concluding Remarks}\label{ConcludingRemarks}
We have studied the tagged-particle statistics of point particles, which collide elastically and move in single file with random-acceleration dynamics and with Langevin dynamics.  

Sect. \ref{sect-homogeneousdist} is concerned with tagged particle statistics in a system of an infinite number of particles, with homogeneous density and with Gaussian-distributed initial velocities.  Both random and equally spaced initial positions are considered. A well-known property of single-file motion with Brownian dynamics is the anomalously slow or subdiffusive growth of the mean-square displacement with time. For random-accelertion and Langevin dynamics, we find the same mean square displacement as for Brownian dynamics (see (\ref{msd-Brownian-annealed}) and (\ref{msd-Brownian-quenched})), except that $2Dt$ is replaced by $\sigma_{v_0}t^2+{2\over 3}\gamma t^3$ and $\sigma_{\rm Lan}^2$, defined in (\ref{sigmaLangevin}), respectively.

These findings are in agreement with the heuristic prediction of Percus \cite{Percus10} that, independent of the particular dynamics, the mean-square displacements with and without the no-passing restriction satisfy
\begin{equation}
\left\langle\left(x_n-x_{n0}\right)^2\right\rangle_{\rm single\;file}\propto \rho^{-1}\left\langle\left(x_n-x_{n0}\right)^2\right\rangle_{\rm non-interacting}^{1/2}\,.\label{Percus}
\end{equation}
Our approach, which utilizes a mapping onto the established Brownian results, does involve details of the dynamics and the initial conditions, but is simple and exact and predicts the proportionality constant in (\ref{Percus}).

For the system of an infinite number of particles with homogeneous initial conditions, we find in Sect. \ref{sect-homogeneousdist} and in the Appendix that the mean square velocity $\langle v_n^2\rangle$, shown in (\ref{avnandvnsqGauss}) and (\ref{avnandvnsqLangevin}) for random-acceleration and Langevin dynamics, respectively, is unaffected by the no-passing restriction and for arbitrary $t$ is the same as for non-interacting particles. Without explicitly calculating the mean square deviation from the initial velocity $\langle\left(v_n-v_{n0}\right)^2\rangle$, we argue that it has the same large-$t$ behavior as for non-interacting particles.

In Sect. \ref{sect-compactinitialcluster} tagged-particle statistics is studied in the spreading, through single-file motion, of a compact cluster of particles, with all of the particles initially at the origin and with initial velocities that are Gaussian-distributed.  The particles tend to migrate to regions of lower density, and, in contrast to the case of non-interacting particles,  $\langle x_n\rangle$ and  $\langle v_n\rangle$ do not vanish. For random-acceleration and Langevin dynamics,  $\langle x_n\rangle$ and $\langle(x_n-x_{n0})^2\rangle$ are the same as in (\ref{<x_N>-Brownian})-(\ref{nudef}) for Brownian particles, except that $2Dt$ is replaced by $\sigma_{v_0}t^2+{2\over 3}\gamma t^3$ and by $\sigma_{\rm Lan}^2$, defined in (\ref{sigmaLangevin}), respectively. In contrast to the case of homogeneous density, considered in Sect. \ref{sect-homogeneousdist}, the no-passing restriction does not lead to anomalously slow growth. For example, for non-interacting randomly accelerated particles initially at the origin, $\langle(x_n-x_{n0})^2\rangle= \sigma_{v_0}t^2+{2\over 3}\gamma t^3$. On imposing the single-file restriction, this is replaced by $\langle(x_n-x_{n0})^2\rangle=A_n\left(\sigma_{v_0}t^2+{2\over 3}\gamma t^3\right)$, with an amplitude $A_n$, given in (\ref{<xn>-Brownian})-(\ref{nudef}), which is smallest for the middle particle and increases monotonically on proceeding from the middle particle toward either end of the file.

The case of initial positions and velocities which are both Gaussian distributed, with zero mean and standard deviation $\sigma_{x_0}$ and $\sigma_{v_0}$, respectively, is considered in the Appendix. In the limit $N=\sqrt{2\pi}\,\rho_0\,\sigma_{x_0}\to\infty$ with constant $\rho_0$, the particles are spread along the $x$ axis with homogeneous density as in Sect. \ref{sect-homogeneousdist}. In this limit, $\langle v_n^2\rangle$, for random-acceleration and Langevin dynamics, is shown to be the same as for non-interacting particles. In the limit                                                                                                                                                                                                                                                                                                                                                  $\sigma_{x_0}\to 0$ with fixed $N$, all of the particles are initially at the origin, as in Sect. \ref{sect-compactinitialcluster}. 

Finally we note that single-file statistics with Langevin dynamics has been studied by Taloni and Lomholt \cite{TaloniLomholt}, but with emphasis on different quantities than the equal-time averages involving  $x_n$ and $v_n$ considered here.\\

\noindent {\bf Acknowledgements}\\
\noindent I thank Ahmed Fouad and Edward Gawlinski for discussions about single-file diffusion and for sharing the results of their simulations.  
\appendix
\section*{Appendix: Analysis of $\langle v_n^2\rangle$ for random-acceleration and Langevin dynamics}   
Averaged over Gaussian distributions of initial positions and initial velocities with mean values zero and standard deviations $\sigma_{x_0}$ and 
$\sigma_{v_0}$, respectively, the $N$-particle probability density (\ref{propN-ranacc1}) for single-file motion with random-acceleration dynamics takes the form 
\begin{eqnarray}
&&\hat{P}_N(\vec{x},\vec{v};t)=N!\prod_{i=1}^N {\sqrt{ab}\over\pi}\,e^{-a(v_i-cx_i)^2-bx_i^2}\,,\label{propNbar-ranacc}\\[2mm]
&&\quad a={1\over 2}\,{\sigma_{x_0}^2+ \sigma_{v_0}^2t^2+\textstyle{2\over 3}\gamma t^3 \over \sigma_{x_0}^2(\sigma_{v_0}^2+2\gamma t)+\gamma t({2\over 3}\sigma_{v_0}^2t^2+{1\over 3}\gamma t^3)}\,,\label{propNbar-ranacc-a}\\
&&\quad b={1\over 2}\,(\sigma_{x_0}^2+ \sigma_{v_0}^2 t^2+\textstyle{2\over 3}\gamma t^3)^{-1}\,,\label{propNbar-ranacc-b}\\
&&\quad c=t(\sigma_{v_0}^2+\gamma t)(\sigma_{x_0}^2+ \sigma_{v_0}^2t^2+\textstyle{2\over 3}\gamma t^3)^{-1}\,.\label{propNbar-ranacc-c}
\end{eqnarray}
The probability density $\hat{P}^{(n)}(x_n,v_n;t)$ of particle $n$, is obtained by integrating (\ref{propNbar-ranacc}) over all of the $x_i$  except $x_n$, with constraint $-\infty<x_1<x_2<\dots <x_N<\infty$, and all of the $v_i$ except $v_n$ from $-\infty$ to $\infty$. This yields
 \begin{eqnarray}
&&\hat{P}^{(n)}(x,v;t)={N!\over(n-1)!(N-n)!}{\sqrt{ab}\over\pi}\,e^{-a(v-cx)^2-bx^2} \hat{f}(x)^{n-1}\hat{f}(-x)^{N-n} \,,\label{tag-dist-ranacc}\\
&&\hat{f}(x)=\sqrt{{b\over\pi}}\int_{-\infty}^x e^{-bx'^{\, 2}}dx'=\textstyle{1\over 2}\,\left[1+{\rm erf}(\sqrt{b}\,x)\right]\,,\label{tagged-f(x)-ranacc}
\end{eqnarray}
where we omit the subscripts on $x_n$ and $v_n$.

The mean number density $\rho(x)$ of  the $N$ particles with Gaussian-distributed initial positions, normalized so that $\int_{-\infty}^\infty\rho(x)dx=N$, is
\begin{equation}
\rho(x)=\rho_0 e^{-x^2/2\sigma_{x_0}^2}\,,\quad \rho_0={N\over\sqrt{2\pi}\,\sigma_{x_0}}\,.\label{Gaussiannumber density}
\end{equation}
In the limit $\sigma_{x_0}\to 0$, the $N$ particles are all initially at the origin, and the tagged-particle distribution (\ref{tag-dist-ranacc}) reduces to the distribution (\ref{aslangul-ranacc-P1bar}). Here we consider the limit $N=\sqrt{2\pi}\,\rho_0\,\sigma_{x_0}\to\infty$ at constant $\rho_0$, corresponding to an infinite number of particles spread homogeneously along the $x$ axis with density $\rho_0$. 

First we focus on particle $n={1\over 2}(N+1)$, i.e., the middle particle in a file of an odd number $N$ of particles. Its mean velocity
$\langle v_{\rm mid}\rangle$ vanishes by symmetry, and, according to  (\ref{tag-dist-ranacc}),   
\begin{eqnarray}
&&\langle v_{\rm mid}^2\rangle={1\over 2a}+c^2\,\langle x_{\rm mid}^2\rangle\,,
\label{vmidsq-1}\\
&&\langle x_{\rm mid}^2\rangle={1\over 2^{N-1}}{N!\over\left({N-1\over2}!\right)^2}\sqrt{{b\over\pi}}\int_{-\infty}^\infty x^2e^{-b x^2}\left[1-{\rm erf}^2(\sqrt{b}\,x)\right]^{{1\over 2}(N-1)}dx\,.\label{xmidsq-1}
\end{eqnarray}

To obtain the mean square velocity for $N=\sqrt{2\pi}\,\rho_0\,\sigma_{x_0}\gg 1$ from (\ref{vmidsq-1}), (\ref{xmidsq-1}), we make use of  the asymptotic forms 
\begin{equation}
a\approx {1\over 2(\sigma_{v_0}^2+2\gamma t)}\,,\quad b\approx{\pi \rho_0^2\over N^2}\,,\quad c\approx{2\pi \rho_0^2\over N^2}\,t\left( \sigma_{v_0}^2+\gamma t\right)\,,
\end{equation}
which imply
\begin{equation}
e^{-b x^2}\left[1-{\rm erf}^2(\sqrt{b}\,x)\right]^{{1\over 2}(N-1)}\approx e^{-2\rho_0^2 N^{-1}x^2}\,.\label{productoffs}
\end{equation}
Thus, 
$\langle x_{\rm mid}^2\rangle\approx {1\over 4}N\rho_0^{-2}$ and $c^2\langle x_{\rm mid}^2\rangle\approx\pi^2 N^{-3}\rho_0^2\,t ^2\left( \sigma_{v_0}^2+\gamma t\right)^2$, and in the limit $N\to\infty$, $\langle v_{\rm mid}^2\rangle\to (2a)^{-1}\to \sigma_{v_0}^2+2\gamma t$. This result is not limited to the middle particle but also holds for all particles a finite distance from it. In this way we are led to $\langle v_n^2\rangle=\sigma_{v_0}^2+2\gamma t$, which confirms (\ref{avnandvnsqGauss}) and is the same as for non-interacting randomly-accelerated particles. 

For Langevin dynamics the results are similar. Equations (\ref{propNbar-ranacc}) and (\ref{tag-dist-ranacc})-(\ref{xmidsq-1}) continue to hold, but with 
\begin{eqnarray}
&& a={1\over 2}\,{\sigma_{x_0}^2+r^2\sigma_{v_0}^2+F\over \left(s^2\sigma_{v_0}^2+G\right)\sigma_{x_0}^2+(s^2 F+r^2 G-2r s H)\sigma_{v_0}^2+FG-H^2}\,,\label{propNbar-Langevin-a}\\
&& b={1\over 2}\,\left(\sigma_{x_0}^2+r^2\sigma_{v_0}^2+F\right)^{-1}={1\over 2}\,\left(\sigma_{x_0}^2+\sigma_{_{\rm Lan}}^2\right)^{-1}\,,\label{propNbar-Langevin-b}\\
&&c={r s\, \sigma_{v_0}^2+H\over\sigma_{x_0}^2+r^2\sigma_{v_0}^2+F}\,.\label{propNbar-Langevin-c} 
\end{eqnarray}
Here $r=\lambda^{-1}\left(1-e^{-\lambda t}\right)$, $s=e^{-\lambda t}$, and the quantities $F$, $G$, $H$, $\sigma_{_{\rm Lan}}^2$ are defined in (\ref{prop1-Langevin}) and (\ref{sigmaLangevin}).
For $N=\sqrt{2\pi}\,\rho_0\,\sigma_{x_0}\gg 1$,
\begin{equation}
a\approx{1\over 2\left(s^2\sigma_{v_0}^2+G\right)}\,,\quad b\approx{\pi\rho_0^2\over N^2}\,,\quad c\approx{2\pi\rho_0^2\over N^2}\left(r s\, \sigma_{v_0}^2+H\right)\,.
\end{equation}
Thus, in the limit $N\to\infty$, $\langle v_{\rm mid}^2\rangle\to (2a)^{-1}\to s^2\sigma_{v_0}^2+G$, which agrees with (\ref{avnandvnsqLangevin}) and is the same as for non-interacting particles with Langevin dynamics.

\begin{figure}
\begin{center}
\includegraphics[width=0.8\textwidth]{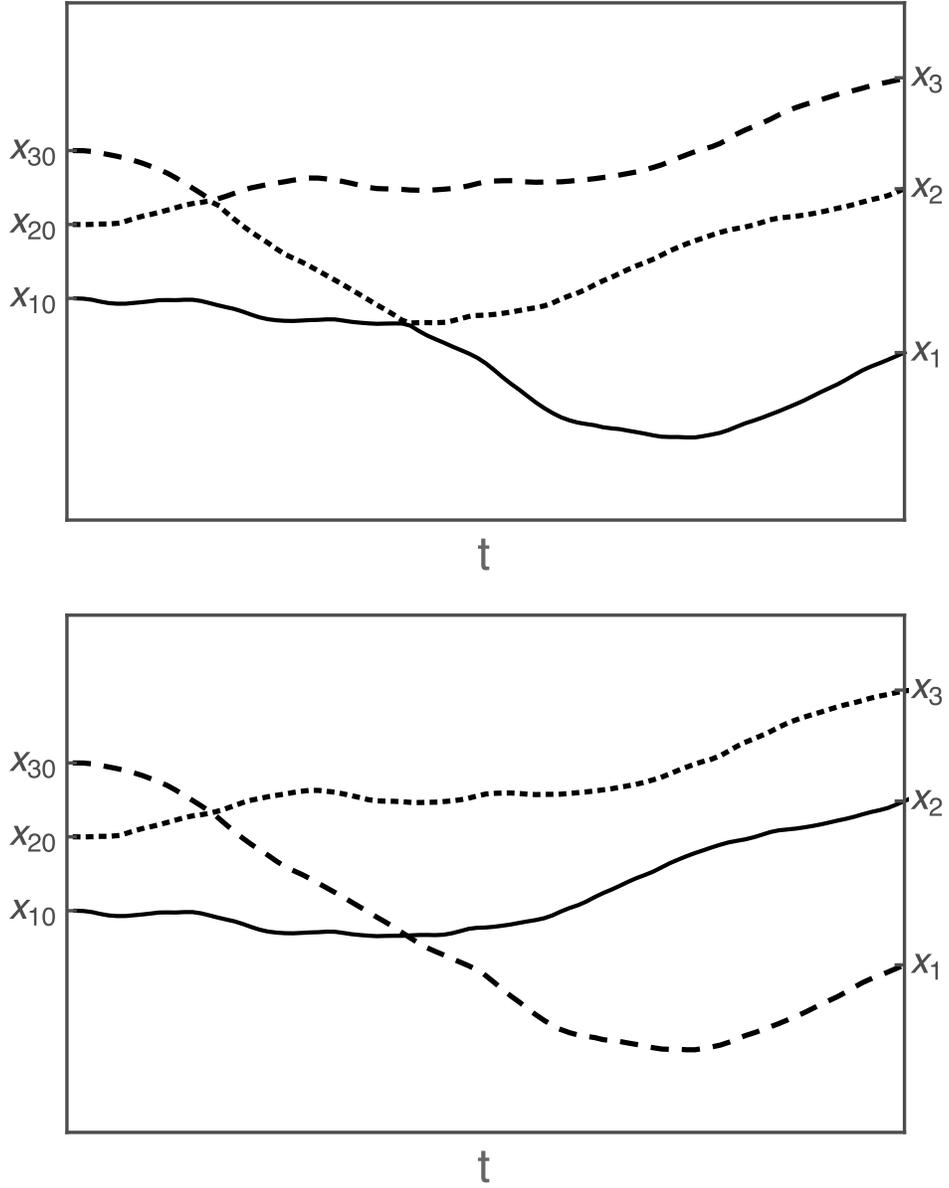}
\caption{The solid, dashed, and dotted curves may be interpreted as the trajectories of three particles undergoing single-file motion (upper graph) or as the trajectories of three non-interacting particles, which are free to pass each other (lower graph).} 
\end{center}
\end{figure}

\begin{figure}
\begin{center}
\includegraphics[width=0.8\textwidth]{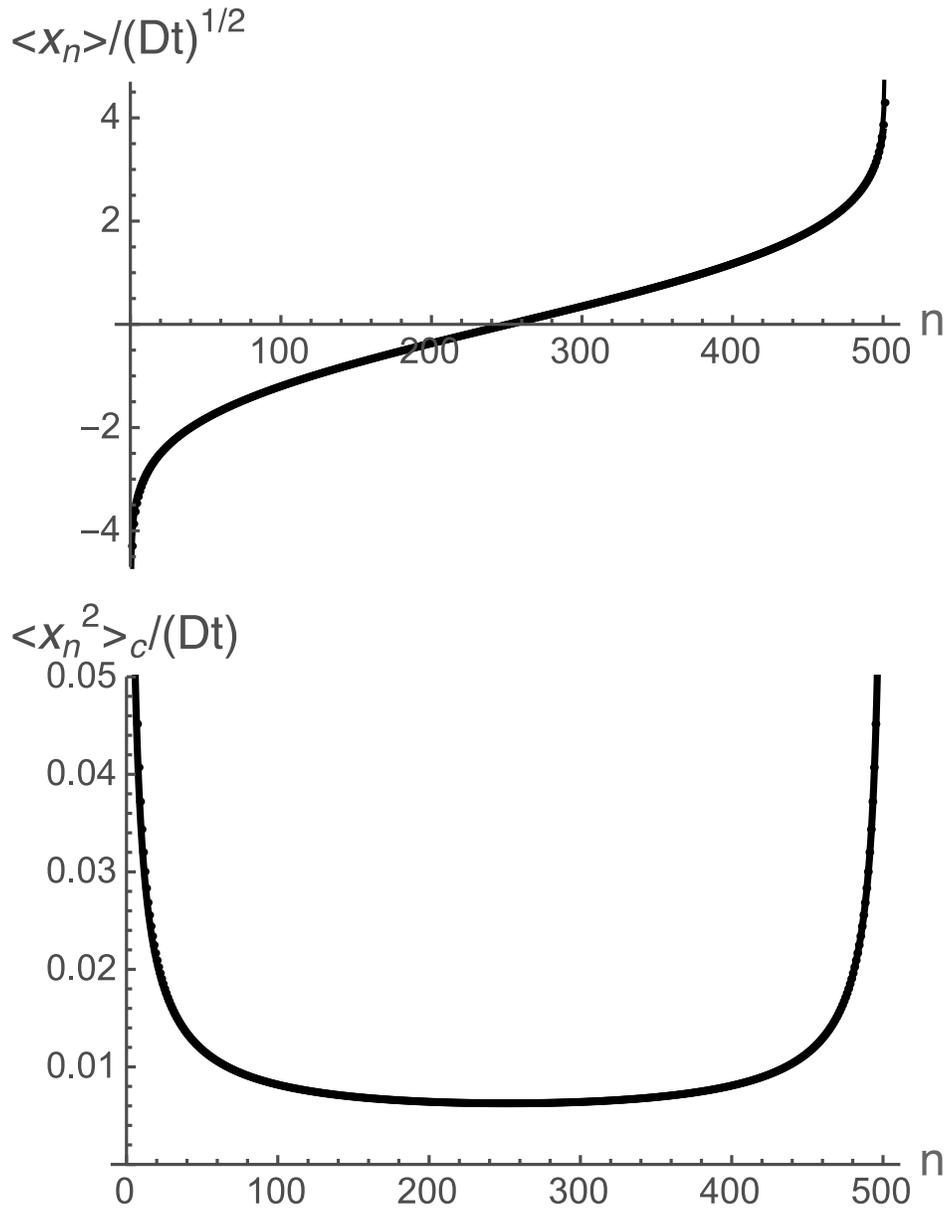}
\caption{Mean displacement $\langle x_n\rangle$ of particle $n$ and its variance $\langle x_n^2\rangle_c=\left\langle\left(x_n-\langle x_n\rangle\right)^2\right\rangle$ in the single-file motion of a system of $N=501$ Brownian particles, all of which are initially at the origin. Three quantities are compared on each graph: the essentially exact prediction from integrating (\ref{propN-Brownian3}) numerically (black points) and the steepest descent result (\ref{<xn>-Brownian})-(\ref{nudef}) for $\alpha=1$ (lower curve) and for $\alpha=-1$ (upper curve). On the scale of the graph the three sets of results appear as a single curve.} 
\end{center}
\end{figure}

\begin{figure}
\begin{center}
\includegraphics[width=0.8\textwidth]{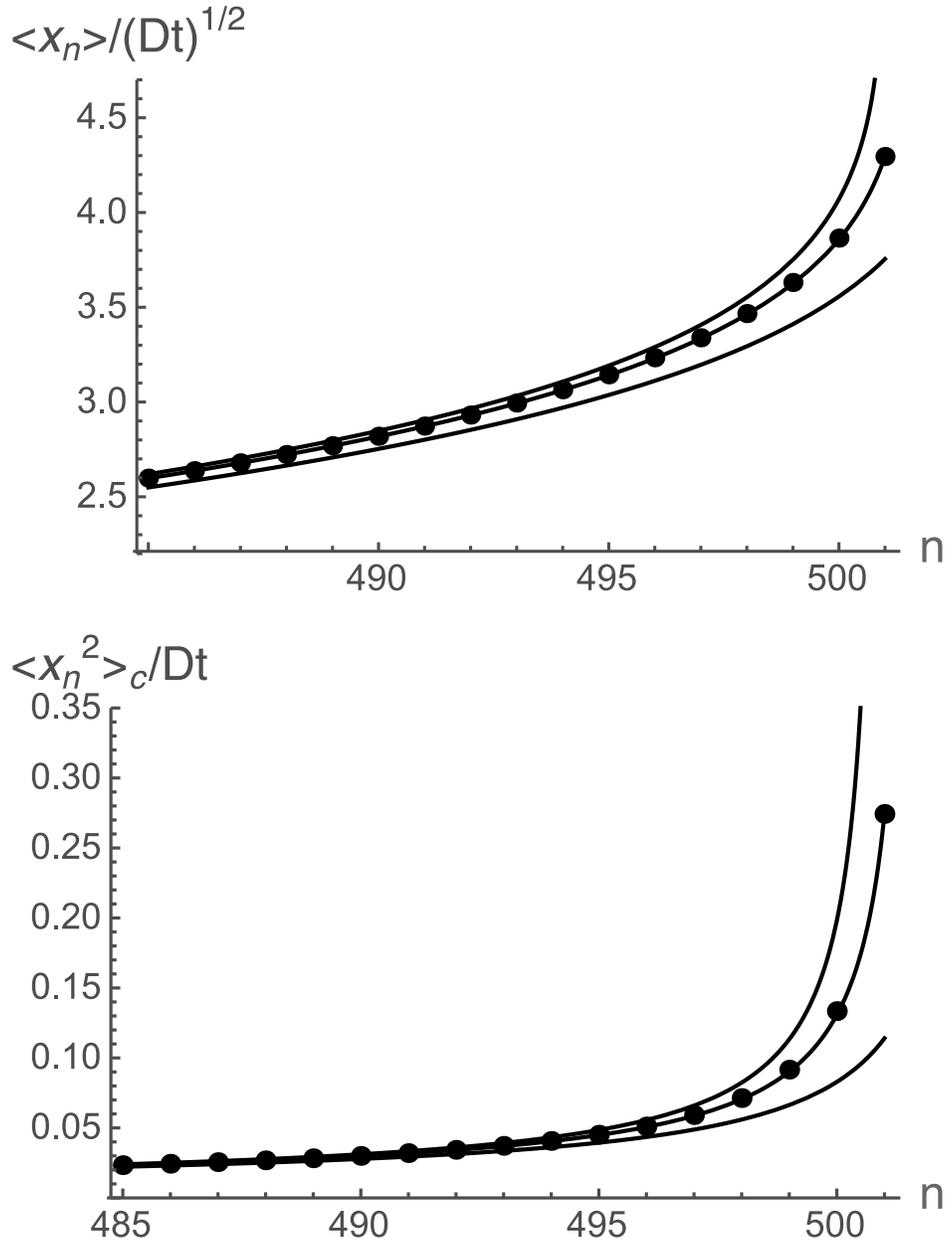}
\caption{Same as Fig. 2, but plotted on an expanded scale, to show the difference in results near the end of the file of $N=501$ particles. Four quantities are compared on each graph: the essentially exact prediction from integrating (\ref{propN-Brownian3}) numerically (black points), and the steepest-descent result (\ref{<xn>-Brownian})-(\ref{nudef}) for $\alpha=1$ (bottom curve), for $\alpha=-1$ (top curve), and for the value $\alpha^*$ (middle curve) which exactly reproduces the rightmost black point and gives an excellent fit for all 501 particles. For the upper and lower graphs  $\alpha^*=-0.401816$ and $-0.324191$, respectively.} 
\end{center}
\end{figure}

\end{document}